\documentclass[12pt,a4paper]{article}
\usepackage[cp866]{inputenc}
\usepackage{amsmath}
\usepackage{axodraw}
\usepackage[dvips]{epsfig}
\usepackage{graphicx}
\usepackage{epsfig}

\newcommand{\vk}{\vec{k}}

\newcommand{\vks}{\vec{k}^{\;2}}
\newcommand{\q}{\vec{q}}
\newcommand{\qs}{\vec{q}^{\;2}}

\newcommand{\x}{\vec{r}}
\newcommand{\xs}{\vec{r}^{\;2}}

\newcommand{\ec}{\vec{e}^{\;*}}
\def\pnot{\mbox{${\not{\hbox{\kern-3.0pt$p$}}}$}}
\def\qnot{\mbox{${\not{\hbox{\kern-2.0pt$q$}}}$}}
\def\enot{\mbox{${\not{\hbox{\kern-2.0pt$e$}}}$}}
\def\knot{\mbox{${\not{\hbox{\kern-2.0pt$k$}}}$}}

\newcommand{\be}{\begin{equation}}
\newcommand{\ee}{\end{equation}}
\newcommand{\ben}{\begin{equation*}}
\newcommand{\een}{\end{equation*}}
\newcommand{\ar}{\begin{array}}
\newcommand{\arn}{\end{array}}
\begin{document}
\numberwithin{equation}{section}     
\sloppy                              
\renewcommand{\baselinestretch}{1.0} 

\begin{titlepage}

\vskip 3cm
\begin{center}
{\bf  Impact Factors for Reggeon-Gluon Transitions} 
\end{center}

\centerline{V.S.~Fadin}
\vskip .6cm
\centerline{\sl 
Budker Institute of Nuclear Physics of SD RAS, 630090 Novosibirsk
Russia}
\centerline{\sl and Novosibirsk State University, 630090 Novosibirsk, Russia}
\vskip 2cm

\begin{abstract}
General expressions for the impact factors up to terms vanishing at the space-time dimension $D\rightarrow 4$ are presented.  Their infrared behaviour is analysed and calculation  of exact in  $D\rightarrow 4$  asymptotics at small momenta of Reggeized gluons is discussed.  
\end{abstract}
  
\vskip 7cm 
\vfill \hrule \vskip.3cm \noindent $^{\ast}${\it Work supported 
 by the Ministry of Education and Science of the Russian Federation and by the RFBR grant 13-02-01023.} 

\vfill $
^{\dag}\mbox{{\it e-mail address:}} 
\mbox{fadin@inp.nsk.su}
$
\end{titlepage}

\section{Introduction}
Reggeon-gluon impact factors  are natural  generalization of particle-particle ones.  In the BFKL approach \cite{Fadin:1975cb}-\cite{Balitsky:1978ic}, discontinuities of elastic amplitudes are given by the convolutions of  the Green functions  of  two interacting Reggeized gluons with the impact factors  of colliding particles describing scattering of these  particles   due to interaction with the Reggeized gluons. Similarly, discontinuities  of many-gluon amplitudes in the multi-Regge kinematics (MRK) contain the Reggeon-gluon impact factors, which  describe transitions of Reggeons (Reggeized gluons) into particles (ordinary gluons) due to interaction with the Reggeized gluons.   These impact factors   appeared  firstly \cite{Bartels:2003jq} in the derivation of the bootstrap conditions for the gluon Reggeization (more precisely, for the  multi-Regge form of the  many-gluon amplitudes). The  idea  of this form  is  the basis of the BFKL approach. It can be  proved  using  the $s$-channel unitarity.
Compatibility of the   unitarity with the multi-Regge form  leads to  the bootstrap relations connecting  discontinuities of the amplitudes with products of their real parts and gluon trajectories \cite{Fadin:2006bj}.  It turns out  that   fulfilment of an infinite set of these relations guarantees the multi--Regge form of scattering  amplitudes. On the other hand,  all bootstrap relations  are fulfilled if several conditions imposed on the Reggeon vertices and the trajectory (bootstrap conditions) hold true \cite{Fadin:2006bj}. Now fulfilment of all bootstrap conditions is proved.  The most complicated condition, which includes the impact factors for Reggeon-gluon transition, was
proved  recently,  both in QCD \cite{Kozlov:2011zza}-\cite{Kozlov:2012zza} and  in  supersymmetric Yang-Mills theories \cite{Kozlov:2014gaa}. 

Discontinuities of $n$-gluon amplitudes in the MRK  at $n \ge 6$ can be used \cite{Fadin:2014yxa} for a simple demonstration of violation of the  ABDK-BDS (Anastasiou-Bern-Dixon-Kosower ---  Bern-Dixon-Smirnov) ansatz \cite{Anastasiou:2003kj, Bern:2005iz} for amplitudes  with maximal helicity violation (MHV) in Yang-Mills theories with maximal supersymmetry (N=4 SYM) in the planar limit and for the calculations of the remainder functions to this ansatz. There are two hypothesis about the remainder functions: the hypothesis of the dual conformal invariance \cite{Bern:2006ew}-\cite{Nguyen:2007ya}, which asserts that the MHV amplitudes are given by the products of the BDS amplitudes and the remainder functions depending only on the anharmonic ratios of kinematic invariants, and the hypothesis of scattering amplitude/Wilson loop
correspondence  \cite{Drummond:2007aua, Drummond:2007cf}, \cite{Brandhuber:2007yx}-\cite{Drummond:2008aq}, according to which  the remainder functions are given by the expectation values of the  Wilson loops.  Both these hypothesis are not proved. They can be tested by comparison of the BFKL discontinuities with the
discontinuities calculated with their use \cite{Lipatov:2010qg}-\cite{Fadin:2011we}.

The discontinuities of many-particle amplitudes are interesting also because they are necessary for  further development of the BFKL approach.  They  do not need  for derivation of the BFKL equation in the next-to-leading logarithmic approximation (NLLA), because they are  suppressed by one power of some of large logarithms  in comparison with the real parts of the amplitudes and therefore  in the NLLA they  don't contribute in the unitarity relations.  But  their   account in the next-to-next-to-leading logarithmic approximation (NNLLA)  is  indispensable.  

All this makes  calculation of  discontinuities of the MRK  amplitudes to be  very important. Since the discontinuities contain the Reggeon-gluon impact factors, the calculation requires  knowledge  of these impact factors  and investigation of their properties  very important.  Here I discuss the current situation with the  Reggeon-gluon impact factors. 

\section{Reggeon-gluon impact factors in the bootstrap scheme}
As it is known, in the next-to-leading order (NLO)  impact factors are scheme dependent. In the Yang-Mills theories of general form they contain contributions of gauge bosons (gluons), fermions and scalars. In the scheme adapted  for  verification of the bootstrap conditions (bootstrap scheme) these contributions were calculated in \cite{Kozlov:2012zz}, \cite{Kozlov:2011zza} and  \cite{Kozlov:2014gaa} respectively. Using these results, one can obtain in these scheme the NLO  Reggeon-gluon impact factors in the Yang-Mills theories with fermions and scalars in any representations of the colour group.
 
Here,  the notation of Refs.\cite{Kozlov:2011zza}-\cite{Kozlov:2014gaa} are  used, in particular, the momentum expansion $p=p^+n_1+p^-n_2+p_\perp$, where $n_{1,2}$~ are the  light-cone vectors, $(n_1,n_2)=1$, \; and $\bot$ means transverse to the $n_1,\,n_2$ plane components.   
For amplitudes with the negative signature, the impact factor  of the transition of the Reggeon $R$  into the gluon $G$  in the interaction with the Reggeized gluons  ${\cal G}_1$ and ${\cal G}_2$ is  antisymmetric with respect to the  ${\cal G}_1\leftrightarrow {\cal G}_2$ exchange. It can be written  as the difference of the $s$ and $u$ parts   
\be 
\langle G R_1| =\langle G R_1|_s - \langle G R_1|_u~, \;\; 
\langle G R|{\cal G}_1{\cal G}_2\rangle_u = \langle G R|{\cal G}_2{\cal G}_1\rangle_s~. \label{s - u}
\ee 
In the NLO each of the parts  contains two  colour structures. In the light-cone gauge $(e(k),n_2)=0, \;\; $ 
\be
e=e_\perp -\frac{(e_\perp k_\perp)}{k^+}n_2
\ee
for the gluon $G$ with the momentum 
$k$ and the polarization vector $(e(k)$,  the $s$ -part  has the form 
\[
\langle G  R_1|{\cal G}_{1}{\cal G}_{2}\rangle_s
= g^2\delta(\q_1-\vk-\x_1-\x_2)\ec \Biggl[ \left(T^{a}T^{b}\right)_{c_1c_2 }\; \Biggl(2\vec{C}_1 + \bar{g}^2 \vec{\Phi}_1(\q_1, \vk; \x_1, \x_2) \Biggr)
\]
\be
+\frac{1}{N_c} Tr\left(T^{c_2}T^{a}T^{c_1}T^{b}\right)\; \bar{g}^2 \vec{\Phi}_2(\q_1, \vk; \x_1, \x_2) \Biggr]~. \label{s part representation}
\ee
Here $g$ is the bare  coupling constant, $\bar g^2=g^2\Gamma(1-\epsilon)/(4\pi)^{2+\epsilon}$, $\Gamma(x)$ is the Euler gamma-function, $\epsilon= (D-4)/2$, $D$ is the space-time dimension, $T^{i}$ are the colour group generators in the adjoint representation,  $q_1, \; k, \; r_1,\; r_2\; $ and $a, \; b\,\; c_1,\; c_2\;   $ are the momenta and colour indices of the Reggeon $R_1$, the gluon $G$ and the Reggeized gluons ${\cal G}_{1}$ and ${\cal G}_{2}$ respectively,  the vector sign is used for  transverse components of vectors,  
\[
\vec{C}_1 \ = \ \q_1 - (\q_1-\x_1)\frac{\qs_1}{(\q_1-\x_1)^2}.
\]
In the bootstrap scheme with the dimensional regularization 
\[
\vec{\Phi}_1(\q_1, \vk; \x_1, \x_2)_{*}=
\vec{C}_1\Biggl(\ln\left(\frac{(\q_1-\x_1)^2}
 {\vks}\right)\ln\left(\frac{\xs_2}{\vks} \right)
 +\ln\left(\frac{(\q_1-\x_1)^2\qs_1}{\vk^{\;4}}\right)\ln\left(\frac{\xs_1}{\qs_1} \right) 
\]
\[ 
 -4\frac{(\vks)^\epsilon}{\epsilon^2}+6\zeta(2)\Biggr)
+\vec{C}_2 
\Biggl(\ln\left(\frac{\vks}{\xs_2}
\right)\ln\left(\frac{(\q_1-\x_1)^2}{\xs_2}\right)
+\ln\left(\frac{\qs_2}{\qs_1}\right)\ln\left(\frac{\vks}{\qs_2}\right)\Biggr)
\]
\[
+2\Big[\vec{C}_1\times\Bigl[\q_1\times \x_1\bigr]\Big]I_{\q_1,  -\x_1}
+2\Big[\vec{C}_2\times\Bigl[\q_1\times \vk\bigr]\Big]I_{\q_1,  -\vk}
-2\Big[\left(\vec{C}_1-\vec{C}_2\right)\times \Bigl[\vk\times \x_2\Bigr]\Big]I_{\vk, \x_2}~.
\]
\[
+\frac{\beta_0}{N_c}\Bigl[\vec C_2 \ln\bigl(\frac{\qs_2(\q_1-\x_1)^2}{\qs_1\xs_2}\bigr) -\vec C_1\Bigl(\frac{1}{\epsilon} +\ln\bigl(\frac{(\q_1-\x_1)^2\xs_1}{\qs_1}\bigr) \Bigr) \Bigr]+ \vec C_1 \Bigl(\frac{67}{9}-\frac{10a_f}{9}-\frac{4a_s}{9}\Bigr)
\]
\[ 
+\Biggl[\frac{\beta_0}{N_c}\Bigl(\vec C_2  \frac{\qs_1+\qs_2}{\qs_1-\qs_2}
+\frac{\vk}{\vks}\frac{2\qs_1\qs_2}{\qs_1-\qs_2}\Bigr)\ln\bigl(\frac{\qs_1}{\qs_2}\bigr)
+\frac{\tilde \beta_0}{N_c}\Biggl(\biggl(\vec C_2  \frac{2\vks}{(\qs_1-\qs_2)^2}
-\frac{\vk(2\vks-\qs_1-\qs_2)}{(\qs_1-\qs_2)^2}\biggr)
\]
\be
\times \biggl(\qs_1+\qs_2-\frac{2\qs_1\qs_2}{\qs_1-\qs_2}\ln\bigl(\frac{\qs_1}{\qs_2}\bigr)\biggr)
+\frac{\vk}{\qs_1}\Biggr)
-(\q_1\rightarrow \q_1-\x_1, \; \q_2\rightarrow \x_2, \; \vk\rightarrow \vk) \Biggr]~.\label{impact-star}
\ee
Here  the subscript $*$ denotes the bootstrap scheme, $\zeta(n)$ is the Riemann zeta-function ($\zeta(2)=\pi^2/6$),
\be
\vec{C}_2 \ = \ \q_1 - \vk\frac{\qs_1}{\vks}, 
\ee
$\bigl[\vec a \times c \bigl[ \vec b\times \vec c\bigr]\bigr]$ is a double vector product, 
\[
I_{\vec p, \vec q}=
\int_{0}^{1}\frac{dx}{(\vec p +x\vec q)^{2}}\ln\left(\frac{\vec p^{\;2}}
{x^2\vec q^{\;2}}\right)~, \;\; I_{\vec p,\vec q}=I_{-\vec p,-\vec q}=I_{\vec q, \vec p}=I_{\vec p,-\vec p
-\vec q}~,  
\]
\[ 
{\beta_0}=\frac{11}{3}{N_c}-\frac{2}{3}a_f-\frac{1}{6}a_s~, {\tilde \beta_0}=\frac{1}{3}{N_c}-\frac{1}{3}a_f+\frac{1}{6}a_s~,   
\] 
$a_f=2\kappa_f n_fT_f,
\;\;a_s=2\kappa_s n_s T_s$, $T_f$ and $T_s$  are defined by the relations
\begin{equation}
 \ \mbox{Tr}\left(T_f^aT_f^b\right)=T_f\delta^{ab}, \;\;\; \mbox{Tr}\left(T_s^aT_s^b\right)=T_s\delta^{ab},
\end{equation}
where $T_f^a$ and $T_s^a$ are the colour group generators for
fermions and scalars, respectively, and $\kappa_f$ ($\kappa_s$)
is equal to $1/2$ for Majorana fermions (neutral
scalars) in  self-conjugated representations
and $1$ otherwise. In the case of $n_M$ Majorana fermions
and $n_s$ scalars  in the adjoint representation $a_f = n_M N_c, \;\;a_s = n_s N_c$. For $N$-extended SYM $n_M=N, \;\; n_s=2(N-1)$. 
Remind that the result \eqref{impact-star} is obtained  
in the  dimensional  regularization, which differs from the dimensional reduction used in supersymmetric theories.   For $N=4$ SYM in the dimensional reduction one has to take $n_s=6-2\epsilon$. In this case the terms with $\beta_0, \; \tilde \beta_0$ and $\Bigl(\frac{67}{9}-\frac{10a_f}{9}-\frac{4a_s}{9}\Bigr)$ in \eqref{impact-star} disappear. Note that the expression \eqref{impact-star}   is obtained with the accuracy up to  terms  vanishing at $\epsilon\rightarrow 0$.  With the same accuracy 
\[
\vec{\Phi}_2(\q_1, \vk; \x_1, \x_2)_{*}= 
\qs_1\int_0^1 \,dx_1\Biggl\{
\frac{(\q_1-\x_1)}{(\q_1-\x_1)^2}\biggl[
\frac{(\vks)^{\epsilon}}{ x_1^{1-2\epsilon}}
-\zeta_2 +\frac{(\xs_2-x_1\vks)}{(\x_2+x_1\vk)^2}  
\]
\[
\times\ln\Big(\frac{(\x_1+x_2\vk)^2(\q_1-\x_1)^2}{\qs_1 \vks x_2^2}\Big)
-\frac{1}{x_1}\ln\Big(\frac{(\x_1+x_1\vk)^2(\x_2+x_1\vk)^2(\x_2+x_2\vk)^2}{x_2^2\xs_{1}\xs_{2}(\vk+\x_2)^2}\Big)
\biggl]
\]
\[
+\frac{\vk}{\vks}\biggl[\frac{1}{x_1}\ln\Big(\frac{(\x_1+x_1\vk)^2}{\xs_{1}}\Big)+ \frac{x_1\vks}{(\x_2+x_1\vk)^2}
\ln\Big(\frac{(\x_1+x_2\vk)^2(\q_1-\x_1)^2}{\qs_1 \vks x_2^2}\Big)\biggr]
\]
\be
-\frac{\q_{1}}{\qs_{1}}\frac{1}{x_1}
\ln\Big(\frac{(\x_1+x_2\vk)^2(\x_1+x_1\vk)^2}{(\x_1+\vk)^2\xs_{1}}\Big)
\Biggr\}\,. \label{F 2}
\ee
Eqs. \eqref{impact-star}, \eqref{F 2} give the impact factors in the bootstrap scheme. Transition to the standard scheme and to the scheme in which  the BFKL  kernel in $N=4$ SYM  and the energy evolution parameter are  invariant under M\"{o}bius transformations in the momentum space is discussed in \cite{Fadin:2014gra}. 
\section{Colour decomposition} 
To calculate discontinuities one needs to  decompose the colour structures  into  irreducible representations of the colour group in the channel with two Reggeized gluons.  The decomposition looks as follows  
\[
\left(T^{a}T^{b}\right)_{c_1c_2 } = N_c\sum_R
c_R\langle ab|\hat{\cal P}_R| c_1 c_2\rangle,\;
\]
\be 
Tr\left(T^{c_2}T^{a}T^{c_1}T^{b}\right) =  N_c\sum_R
c_R(c_R -\frac{1}{2} )\langle ab|\hat{\cal P}_R| c_1 c_2\rangle,\;
\ee
where $\hat{\cal P}_R$ are the projections operators of the two-Reggeon colour states on the irreducible representations $R$. Explicit form of these operators and the values of the coefficients $c_R$  can be found in \cite{Ioffe:2010zz}. In the limit of large $N_c$ the term in \eqref{s part representation} with the  colour structure 
$Tr\left(T^{c_2}T^{a}T^{c_1}T^{b}\right)$ disappears and with the account of \eqref{s - u} the impact factors take the form 
\[
\langle G  R_1|{\cal G}_{1}{\cal G}_{2}\rangle
= g^2\delta(\q_1-\vk-\x_1-\x_2)\ec \Biggl[ f^{abc}f^{cc_1 c_2}\; \Biggl(2\q_1 -  (\q_1-\x_1)\frac{\qs_1}{(\q_1-\x_1)^2}  
\]
\[
- (\q_1-\x_2)\frac{\qs_1}{(\q_1-\x_2)^2} + \frac{\bar{g}^2}{2} \Bigr( \vec{\Phi}_1(\q_1, \vk; \x_1, \x_2)+\vec{\Phi}_1(\q_1, \vk; \x_2, \x_1)\Bigr)\Biggr)
+ d^{abc}d^{cc_1 c_2}\;\]
\be
\times  \Biggl[\frac{\qs_1(\q_1-\x_2)}{(\q_1-\x_2)^2}-\frac{\qs_1(\q_1-\x_1)}{(\q_1-\x_1)^2}  + \frac{\bar{g}^2}{2} \Bigr( \vec{\Phi}_1(\q_1, \vk; \x_1, \x_2)-\vec{\Phi}_1(\q_1, \vk; \x_2, \x_1)\Bigr)\Biggr)\Biggr]. 
\label{IF}
\ee 
\section{Infrared behaviour of the impact factors} 
As  is clear from the  foregoing,  Eq. \eqref{IF} gives the  impact factor up to terms vanishing  in the limit $\epsilon\rightarrow 0$.  
Unfortunately, using  \eqref{IF} for calculation of discontinuities does not provide such accuracy for them. The reason is the integration measure 
$d^{2+2\epsilon}r_{1\bot}d^{2+2\epsilon}r_{1\bot}/(r_{1\bot}^2r_{2\bot}^2)\delta(q_{2\bot}-r_{1\bot}-r_{2\bot})$ which is singular at $\epsilon\rightarrow 0$. To keep in the discontinuities  all terms nonvanishing   in the limit $\epsilon\rightarrow 0$ one has to calculate $\vec{\Phi}_1(\q_1, \vk; \x_1, \x_2)$ more accurately. 

In fact, greater accuracy is required  only in the region of small $|\x_2|$, because in the limit $|\x_1|\rightarrow 0$ $\vec{\Phi}_1(\q_1, \vk; \x_1, \x_2)$ turns to be zero, which is seen from \eqref{impact-star}. In contrast, in the limit   $|\x_2|\rightarrow 0$  $\vec{\Phi}_1(\q_1, \vk; \x_1, \x_2)$ 
not only does not vanish but have logarithmic singularities. To keep in the  discontinuities all terms nonvanishing   in the limit $\epsilon\rightarrow 0$ 
one has to know in the region of small $|\x_2|$ terms of order  $\epsilon$ in   $\vec{\Phi}_1(\q_1, \vk; \x_1, \x_2)$  and  must not expand $(\xs_2)^\epsilon$ in powers of   $\epsilon$. 

In the NLO, the impact factor contains  contributions of  two types: virtual ones, which are  obtained  from the one-loop corrections to the Reggeon vertices and the gluon trajectory, and real  contributions arising  from production of two real particles. 
In the bootstrap scheme, the real contribution to $\vec{\Phi}_1(\q_1, \vk; \x_1, \x_2)$  can be calculated at small $|\x_2|$  exactly in $\epsilon$ using intermediate results of Refs. \cite{Kozlov:2011zza}-\cite{Kozlov:2014gaa}. 
It is  proportional to $(\xs_2)^\epsilon$ and has the form \cite{Fadin:2015}
\[
\vec{\Phi}_1(\q_1, \vk; \x_1, \x_2)^{real}_* = 4(\xs_2)^\epsilon\frac{\Gamma^2(1+\epsilon)}{\Gamma(1+2\epsilon)}\Biggl[
\vec C_2\Big(\frac{1}{2\epsilon^2}+\frac{(\psi(1)-\psi(1+2\epsilon))}{\epsilon}
\]
\be
-\frac{\Gamma(1+2\epsilon)}{\Gamma(4+2\epsilon)}\Bigl(\frac{a_1(1+\epsilon)}{\epsilon}+ a_2\Bigr)+\frac{2\Gamma(1+2\epsilon)}{\Gamma(4+2\epsilon)}\frac{\x_2(\x_2 \vec C_2)}{\xs_2}a_2\Biggr]~, 
\ee
where 
\be
a_1=11+7\epsilon -2(1+\epsilon)a_f -\frac{a_s}{2}, \;\; a_2=1+\epsilon -a_f +\frac{a_s}{2}~.
\ee
For $N=4$ SYM,  the coefficients $a_1$ and $a_2$ vanish in the dimensional reduction. 
 
The virtual contribution to $\vec{\Phi}_1(\q_1, \vk; \x_1, \x_2)$ in the limit 
of small $|\x_2|$ can be obtained with the required accuracy using its representation \cite{Kozlov:2012zz} in terms of the Reggeon vertices and the gluon  trajectory and exact in $\epsilon$ expressions for the trajectory, the gluon-gluon-Reggeon vertex and fermion and scalar contributions to the Reggeon-Reggeon-gluon vertex which can be found in Refs. \cite{Bartels:2012sw}, \cite{Fadin:1995xg} and \cite{Fadin:1994fj, Kozlov:2014gaa} respectively   and the  gluon production vertex in $N=4$ SYM computed  through to ${\cal O} (\epsilon^2)$ in \cite{DelDuca:2009ae}. Full expressions for the Reggeon-gluon impact factors in the region of small  $|\x_2|$  in the bootstrap and the standard schemes  will be given in  \cite{Fadin:2015}. 

\section{Summary}
The  impact factors for Reggeon-gluon transitions are an integral part  of the BFKL approach. They enter  the expressions for  the discontinuities of many-particle amplitudes and  the bootstrap conditions for the gluon Reggeization,  and enable to demonstrate  in a simple way violation of the  ABDK-BDS ansatz for MHV amplitudes  in N=4 SYM  in the planar limit  and to check the hypotheses about the remainder functions to this ansatz.  Their knowledge is necessary for further development of the BFKL approach. 

Here the impact factors  in Yang-Mills theories with   fermions and scalars in any  representations of the  gauge group  are presented up to terms vanishing at $\epsilon \rightarrow 0$. Their colour decomposition is performed and infrared behaviour is discussed.


\end{document}